\documentclass[fleqn,usenatbib]{mnras}

\usepackage{newtxtext,newtxmath}

\usepackage[T1]{fontenc}

\usepackage{graphicx}	
\usepackage{amsmath}	
\usepackage{subfigure}
\usepackage[british]{babel}

\usepackage{multirow}
\usepackage[normalem]{ulem}

\title[Main perturbing objects on Prometheus and Pandora]{The main perturbing objects on the orbits of (616) Prometheus and (617) Pandora}

\author[A. R. Gomes-Júnior et al.]{
A. R. Gomes-Júnior,$^{1,2}$\thanks{E-mail: altair.gomes@unesp.br, altair.gomes@linea.gov.br}
T. Santana,$^{1}$
O. C. Winter$^{1}$
and R. Sfair$^{1,3}$
\\
$^{1}$UNESP - São Paulo State University, Grupo de Dinâmica Orbital e Planetologia, CEP 12516-410, Guaratinguetá, SP, Brazil\\
$^{2}$Laboratório Interinstitucional de e-Astronomia - LIneA, Rua Gal. José Cristino 77, Rio de Janeiro, RJ 20921-400, Brazil\\
$^{3}$Institut für Astronomie und Astrophysik, Eberhard Karls Universität Tübingen, Auf der Morgenstelle 10, 72076 Tübingen, Germany
}

\date{Accepted XXX. Received YYY; in original form ZZZ}

\pubyear{2021}

\begin{document}
\label{firstpage}
\pagerange{\pageref{firstpage}--\pageref{lastpage}}
\maketitle

\begin{abstract}
The dynamical evolution of the Prometheus and Pandora pair of satellites is chaotic, with a short 3.3 years Lyapunov time. It is known that the anti-alignment of the apses line of Prometheus and Pandora, which occurs every 6.2 years, is a critical configuration that amplifies their chaotic dynamical evolution. However, the mutual interaction between Prometheus and Pandora is not enough to explain the longitudinal lags observed by the Hubble Space Telescope.
The main goal of the current work is to identify the main contributors to the chaotic dynamical evolution of the Prometheus-Pandora pair beyond themselves.
Therefore, in this work, we first explore the sensibility of this dynamical system to understand it numerically and then build numerical experiments to reach our goals.
We identified that almost all major satellites of the Saturn system play a significant role in the evolution of Prometheus' and Pandora's orbits.
\end{abstract}




\section{Introduction}
\label{Sec:introduction}

Observations made during the ring plane crossing of 1995 by the Hubble Space Telescope (HST) revealed that the angular position of Prometheus and Pandora were shifted from the expected values \citep{BoshRivkin96, Nicholson96, McGhee2000}. The accepted explanation for the lags is the chaotic orbital evolution associated with a mean motion resonance between the two satellites \citep{RennerSicardy03, Goldreich2003_chaotic, Goldreich2003_origin}. However, the values of the ratio between the lags of Prometheus and Pandora along the time clearly show that the mutual interaction between the two satellites is far from being enough to explain the measured angular gaps \citep{deSantana2020}.

A few works performed simulations including other satellites. For example, \cite{French2003} determined with high precision the astrometric positions of Prometheus, Pandora, and nine other satellites using a large set of images taken by the Wide Field and Planetary Camera of the HST from 1996 to 2002. By comparing the data from the expected positions, the authors computed mean longitudes and lags for the pair of satellites. They proposed that a resonance involving the F ring could be the common origin of the uncertainty of their positions. 

\cite{Cooper2004} performed numerical simulations of the full equations of motion to study the orbits of Prometheus and Pandora over thirty years under the presence of the eight major satellites of Saturn' system (Mimas, Enceladus, Tethys, Dione, Rhea, Titan, Hyperion, and Iapetus), plus the pair Janus and Epimetheus. The authors confirmed that the anti-correlation in the temporal variation of the two satellites' mean longitudes remains maintained when the others are considered. They also confirmed the chaos on Prometheus orbit by computing its Lyapunov exponent, while for Pandora, it was implied by the sensibility of the initial conditions used on the simulations. Moreover, through theoretical analysis, a role for the pair Janus and Epimetheus through two second-order eccentricities mean motion resonances, 17:15 and 21:19, is also pointed as important for the chaotic behaviour of Prometheus and Pandora, but in a longer timescale. 

In \cite{Renner05} the authors fitted numerical simulations also within the HST data. Using a Radau integrator, the system consisted of the major satellites of the Saturn system: Epimetheus, Janus, Mimas, Enceladus, Tethys, Dione, Rhea, Titan, and Iapetus, and allowed the authors to confirm the chaotic behavior of Prometheus and Pandora. Furthermore, they also derived the masses and densities of the satellites. The authors state that the chaotic interaction between Prometheus and Pandora remains in the addition of the other satellites. They also point out the influence of Mimas through the nearby corotation resonance 3:2 with Pandora, and the Janus and Epimetheus dynamical influence as well, in agreement with \cite{Cooper2004}.
 
The current work aims to identify the main contributors to the chaotic dynamical evolution of the Prometheus-Pandora system beyond themselves. However, the high instability of the system is a major difficulty in reaching our goal. Therefore, in this work, we first explored the sensibility of this dynamical system to understand it numerically and then build up numerical experiments that allow us to identify each of the bodies that significantly contribute to the chaotic behaviour of Prometheus and Pandora. 
 
The work can be divided into three parts. The first one is devoted to better understanding the high sensibility of the Prometheus-Pandora system by comparing our models to the JPL ephemeris.

The second part is concerned with the strong instability generated by the anti-alignment of the apses line of Prometheus and Pandora, which occurs periodically. Any slight orbital difference is significantly amplified during one of these anti-alignments. Consequently, after a few anti-alignments, the angular difference might drastically increase.

The third and last part of this work focuses on identifying Saturn's satellites that are relevant for the chaotic evolution of Prometheus and Pandora, beyond that generated by their mutual interaction (including the effect due to the near 3:2 mean motion resonance between Mimas and Pandora).

Therefore, section \ref{Sec:methodology} describes the approach adopted in the numerical integration of the satellites' orbits. In section \ref{Sec:orbits} we will present the numerical integrations considering twenty objects of the Solar System with different epochs for initial conditions to fit our model to the JPL ephemeris. Section \ref{Sec:antialignment} will numerically verify the anti-alignment implications when studying the orbits of Prometheus and Pandora. Section \ref{Sec:relevant-satellites} will present our numerical experiments, which reveal the more relevant satellites on the dynamical evolution of the Prometheus-Pandora pair. Then, in section \ref{Sec:final-comments} we will provide our final comments.

\section{Methodology}
\label{Sec:methodology}

Using our software, named BOSS (Brazilian Orbital Solution for Satellites), we integrate the equations of motion to determine the dynamical evolution of the orbits of Prometheus and Pandora. The numerical model of Saturn's satellites is in a Saturncentric reference frame with axes in the International Celestial Reference System (ICRS/J2000). The equations of motion are presented in \autoref{Eq:dynamical_model} and described in \cite{Lainey2004_ephem}.

\begin{equation}
\label{Eq:dynamical_model}
\begin{array}{ll}

\ddot{\vec{r_i}}= & \displaystyle -G(M_S+m_i)\frac{\vec{r_i}}{r_i^3} +\sum_{j=1}^{N}Gm_j \left(\frac{\vec{r_j}-\vec{r_i}}{r_{ij}^3} - \frac{\vec{r_j}}{r_j^3} \right)\\
 & \displaystyle +G(M_S+m_i) \nabla U_{\bar{i}\hat{S}} +\sum_{j=1}^N Gm_j\nabla U_{\bar{j}\hat{S}}
\end{array}
\end{equation}
where
\begin{equation}\begin{array}{ll}

U_{\bar{i}\hat{S}}=&\displaystyle -\frac{R_S^2 J_2}{r_i^3}\left(\frac{3}{2}\sin^2 \Phi_i-\frac{1}{2}\right) -\frac{R_S^3 J_3}{r_i^4}\left(\frac{5}{2}\sin^3 \Phi_i - \frac{3}{2}\sin \Phi_i \right) \\ & \\&
\displaystyle-\frac{R_S^4 J_4}{r_i^5}\left(\frac{35}{8}\sin^4 \Phi_i-\frac{15}{4}\sin^2 \Phi_i+\frac{3}{8}\right)\\
& \\
&\displaystyle-\frac{R_S^6 J_6}{r_i^7}\left(\frac{231}{16}\sin^6 \Phi_i-\frac{315}{16}\sin^4 \Phi_i+\frac{105}{16}\sin^2 \Phi_i-\frac{5}{16}\right)

\end{array}
\end{equation}
with the following notation:

\begin{itemize}
    \item $S$ stands for Saturn and corresponding mass $M_S$ and radius $R_S$;
    \item $i$ stands for the satellite whose equations of motion are being considered, its mass $m_i$ and position relative to Saturn $\vec{r_i}$, with respective distance $r_i$;
    \item $j$ corresponds for any of the other $N$ bodies of the Solar System that are perturbing $i$: other satellites, the planets (except Saturn) and the Sun. $m_j$ is the corresponding mass. $\vec{r_j}$ and $r_j$ are the position and distance relative to Saturn respectively;
    \item $r_{ij}$ is the distance between the bodies $i$ and $j$;
    \item $U_{\bar{i}\hat{S}}$ and $U_{\bar{j}\hat{S}}$ are the gravitational potential caused by the oblateness of Saturn experienced by body $i$ and $j$, respectively;
    \item $J_2$, $J_3$, $J_4$ and $J_6$ are zonal harmonic coefficients of Saturn's gravitational potential of 2nd, 3rd, 4th and 6th order, respectively;
    \item $\Phi_i$ is the latitude of the body $i$ in the Saturnian equatorial frame.
\end{itemize}

For the analysis described in Sections \ref{Sec:orbits}-\ref{Sec:relevant-satellites}, different sets of dynamical constants and perturbers will be used. For every case, the values of these constants are the same as the JPL kernel SAT393 \footnote{\url{https://naif.jpl.nasa.gov/pub/naif/generic_kernels/spk/satellites/sat393.cmt}}.  \autoref{tab:constants-sat393} presents the values for the dynamical constants used in our models extracted from SAT393.

\begin{table}
    \centering
    \caption{Dynamical constants from SAT393 used in the model (see \autoref{Sec:methodology})}
    \begin{tabular}{l|c|c}
        Constant & Value & Unit \\ \hline
        Janus GM & 1.265765099012197E-01 & $\textrm{km}^{3}\textrm{s}^{-2}$ \\
        Epimetheus GM & 3.512333288208074E-02 & $\textrm{km}^{3}\textrm{s}^{-2}$ \\
        Helene GM & 3.424829447502984E-04 & $\textrm{km}^{3}\textrm{s}^{-2}$ \\
        Atlas GM & 3.718871247516475E-04 & $\textrm{km}^{3}\textrm{s}^{-2}$ \\
        Prometheus GM & 1.075208001007610E-02 & $\textrm{km}^{3}\textrm{s}^{-2}$ \\
        Pandora GM & 9.290325122028795E-03 & $\textrm{km}^{3}\textrm{s}^{-2}$ \\
        Mimas GM & 2.503629609027271E+00 & $\textrm{km}^{3}\textrm{s}^{-2}$ \\
        Hyperion GM & 3.712505242740757E-01 & $\textrm{km}^{3}\textrm{s}^{-2}$ \\
        Phoebe GM & 5.533878663162022E-01 & $\textrm{km}^{3}\textrm{s}^{-2}$ \\
        Enceladus GM & 7.210841599764501E+00 & $\textrm{km}^{3}\textrm{s}^{-2}$ \\
        Tethys GM & 4.120864966231671E+01 & $\textrm{km}^{3}\textrm{s}^{-2}$ \\
        Dione GM & 7.311562440698799E+01 & $\textrm{km}^{3}\textrm{s}^{-2}$ \\
        Iapetus GM & 1.205075311030973E+02 & $\textrm{km}^{3}\textrm{s}^{-2}$ \\
        Rhea GM & 1.539433320228564E+02 & $\textrm{km}^{3}\textrm{s}^{-2}$ \\
        Titan GM & 8.978137712627313E+03 & $\textrm{km}^{3}\textrm{s}^{-2}$ \\
        Jupiter system GM & 1.267127641000000E+08 & $\textrm{km}^{3}\textrm{s}^{-2}$ \\
        Uranus system GM & 5.794556400000000E+06 & $\textrm{km}^{3}\textrm{s}^{-2}$ \\
        Neptune system GM & 6.836527100580000E+06 & $\textrm{km}^{3}\textrm{s}^{-2}$ \\
        Sun GM & 1.327132332639221E+11 & $\textrm{km}^{3}\textrm{s}^{-2}$ \\
        Saturn GM & 3.793120655618811E+07 & $\textrm{km}^{3}\textrm{s}^{-2}$ \\
        Sat. J$_2$ & 1.629133249525738E-02 &  \\
        Sat. J$_3$ & 1.494723182852077E-06 &  \\
        Sat. J$_4$ & -9.307138534779719E-04 &  \\
        Sat. J$_6$ & 8.943208329411604E-05 &  \\
        Sat. Eq. Radius & 60330 & km \\
        Sat. pole $\alpha_p$ & 40.5839 & degrees \\
        Sat. pole rate $\Dot{\alpha}_p$ & -0.05058 & deg century$^{-1}$ \\
        Sat. pole $\delta_p$ & 83.5377 & degrees \\
        Sat. pole rate $\Dot{\delta}_p$ & -0.00553 & deg century$^{-1}$ \\
        Sat. prime meridian angle & 38.9 & degrees \\
        Sat. rotation rate & 810.79390 & deg day$^{-1}$ \\
        \hline
    \end{tabular}
    \label{tab:constants-sat393}
\end{table}

The equations of motion are integrated using the 15th order integrator RADAU \citep{Everhart1985} with a constant step of 0.025 day. Unless stated otherwise, for each simulation, only the orbits of Prometheus and Pandora are being integrated, while the positions of the perturbing objects come from the ephemeris and are not being redetermined.

\section{On the Orbits of Prometheus and Pandora}
\label{Sec:orbits}

The JPL ephemeris SAT393 was fitted to the observations obtained from 1980 up to 2016. In this period, many observations of Prometheus and Pandora were obtained by Voyager I and II spacecrafts, the Hubble Space Telescope, and, mainly, by the Cassini spacecraft \citep{Cooper2015}.

Our initial goal was to fit our model to the observations. However, a large set of astrometric positions are not publicly available. For instance, the Natural Satellites Database\footnote{\url{http://nsdb.imcce.fr/obspos/obsindhe.htm}} \citep{Arlot2009} of Paris Observatory only contains positions from Cassini up to 2013 and none from the Voyagers. Therefore, determining the astrometry of these satellites from the observations is out of the scope of this work.

Because of that, we will use the ephemeris from the SAT393 kernel to represent the observations. In this regard, we first try to replicate the SAT393 model by using the same constants and initial conditions. In this case, we used all the objects described in \autoref{tab:constants-sat393} in the numerical model. The integration was realised between 1981, near Voyager II's closest approach to Saturn, and 2050, the limit of the SAT393 ephemeris. The positions for all perturbers come from the ephemeris itself.

\begin{figure}
    \centering
    \includegraphics[width=\linewidth]{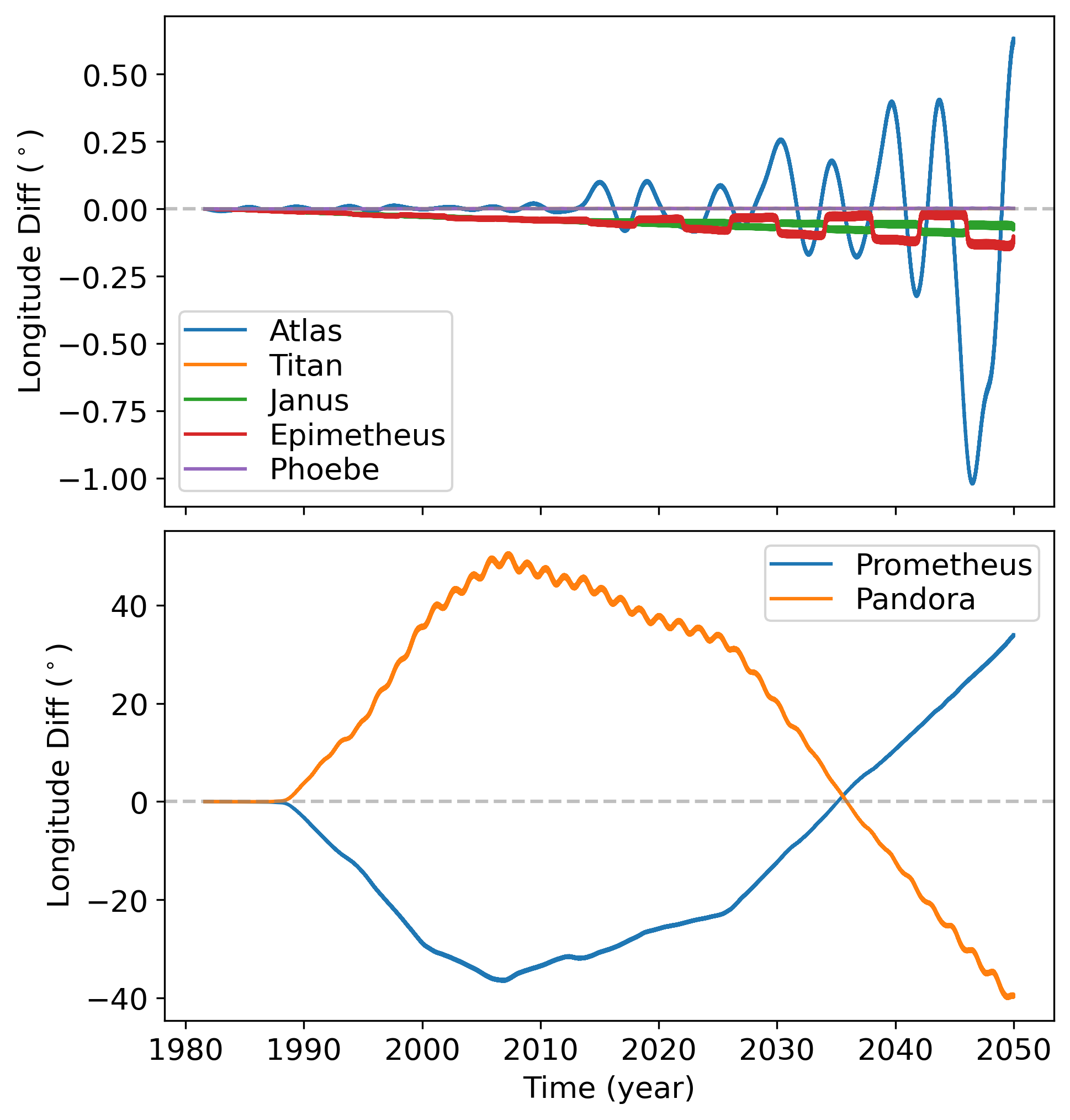}
    \caption{Difference in true longitude between BOSS and JPL SAT393 in integrations where only the referred satellite was integrate and all the perturbers' positions come from the ephemeris.}
    \label{fig:foi_lag_all}
\end{figure}

The same test was applied for other objects of the Saturn system to see their behaviour compared to Prometheus and Pandora: Atlas, Titan, Phoebe, and the pair Janus-Epimetheus. Then, we compared the difference in true longitude between BOSS and SAT393, as shown in \autoref{fig:foi_lag_all}.

The results show that our model retrieved well the orbits of Atlas, Titan, Phoebe, and the pair Janus-Epimetheus. In the period studied, the difference in longitude was smaller than $1\degr$ for Atlas, which is probably due to its 54:53 mean motion resonance with Prometheus, and smaller than $0.2\degr$ for Titan, Janus-Epimetheus, and Phoebe. Up to the last observations by Cassini, in 2016, the difference is smaller than $0.2\degr$ for all satellites.

However, the same test for Prometheus' and Pandora's orbits showed a difference of up to $40\degr$ in true longitude. This difference was reached about 25 years after the beginning of the integration. The more considerable difference observed in Prometheus and Pandora relative to the smaller ones observed with the other satellites confidently tells us this difference is mainly caused by a dynamical interaction rather than a numerical error.

Four other simulations, starting at different epochs (J1985, J1991, J2001 e J2020) and ending in J2050, were also made. These epochs were randomly chosen to verify the behaviour due to different spatial configurations as initial conditions. In all these cases, the difference in true longitude for Prometheus and Pandora between SAT393 and BOSS was still huge, with the maximum difference varying between $20\degr$ and $100\degr$. Furthermore, the behaviour of these differences largely varies depending on the initial epoch.

Finally, we tried to force our model on the orbits of Prometheus and Pandora to be the same as the SAT393 in the period of observations. With this, we expect to have a model indirectly fitted to the \textit{observations} where the parameters can be later changed to study its behaviour. For that, we generated positions of both satellites from the SAT393 ephemeris uniformly distributed in the interval 1980-2020 and used them as \textit{observations} to be fitted in our model. The positions of the perturbing satellites were obtained directly from the SAT393 model, thus their orbits were not being integrated, as described in \autoref{Sec:methodology}. In this case, the initial conditions were from the SAT393 at the epoch J2000, the middle of the chosen interval.

The process of fitting is described in \cite{Lainey2004_fit}. First, the variational equations of the force function relative to the initial conditions are simultaneously integrated with the \autoref{Eq:dynamical_model}. Then, these equations are fitted to the offset between the observations and the numerical model using a Least Squares method, resulting in new initial conditions.

The fit did not converge when we used all the positions directly. To avoid it, we used only a few observations close to the initial epoch, fitted new initial conditions, and then repeated the process, increasing the \textit{observations} for a larger interval of time.

The difficulty of fitting the orbits of Prometheus and Pandora, using all the interval of observations, from Voyager I to Cassini, is a common issue (Lainey, 2020, priv. comm.). Usually, different models are fitted, where each one is suitable for a different interval of time. We were able to fit our model to the SAT393 ephemeris using points distributed over the period 1983 and 2017. Further than that, even with a slight increase in the interval of ephemeris considered, the fit diverges again.

\begin{figure}
    \centering
    \includegraphics[width=\linewidth]{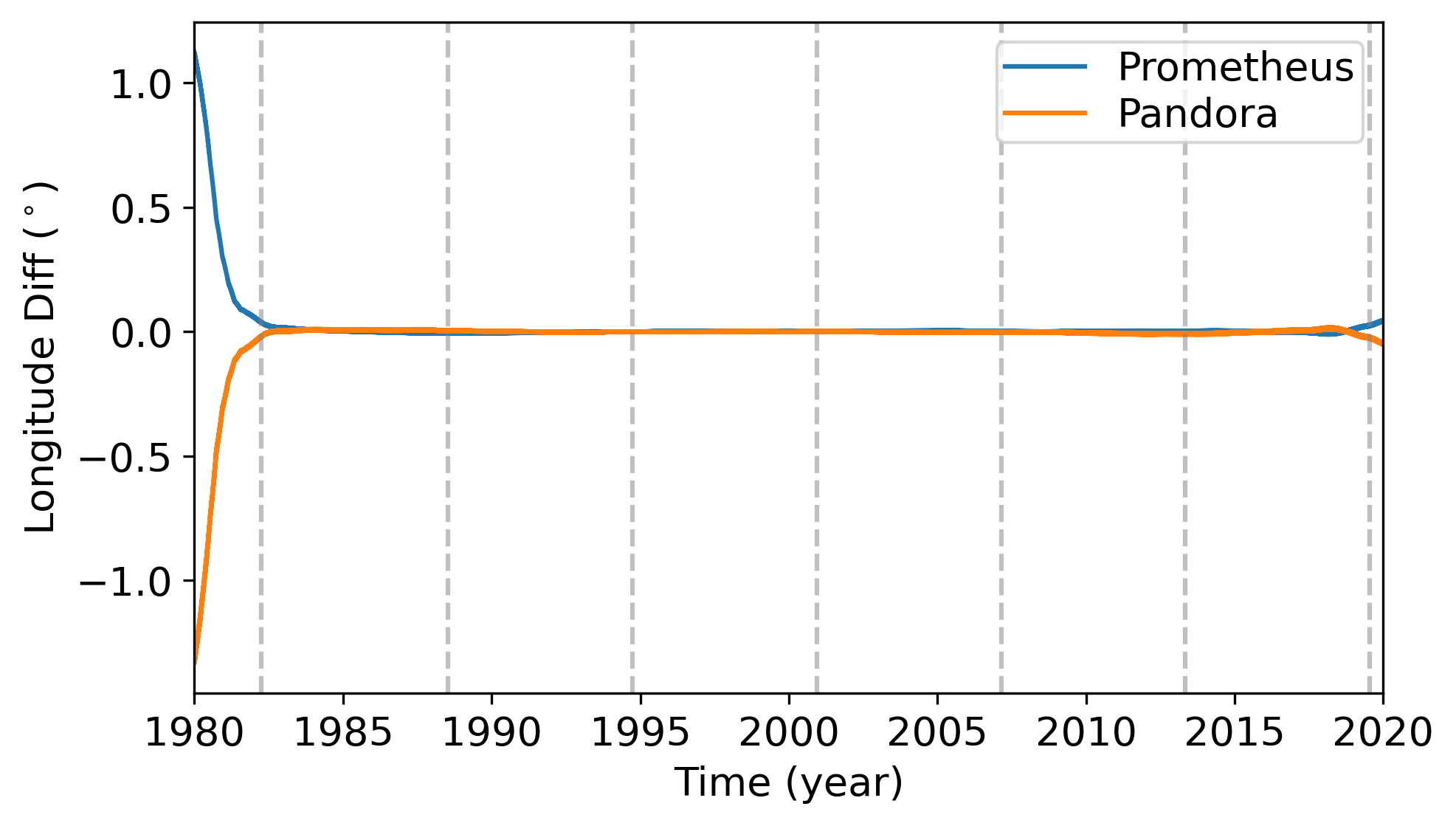}
    \caption{Difference in true longitude between the numerical model obtained from the fitting to the SAT393 ephemeris using BOSS and SAT393 itself. The dashed lines are the anti-alignment between Prometheus and Pandora discussed in \autoref{Sec:antialignment}.}
    \label{fig:fitted_to_jpl}
\end{figure}

\autoref{fig:fitted_to_jpl} shows the difference between BOSS and SAT393 after we have fitted a numerical model to the SAT393 ephemeris between J1983 and J2017. The residuals are mostly on the order of 10 km, with 30 km in the extremities of the interval. The difference in initial conditions are in the order of 1km.

All these tests showed how sensitive the orbits of Prometheus and Pandora are. Unfortunately, their chaotic behaviour prevents us from propagating their trajectories for more than a few years reliably. Therefore, observations must be obtained more frequently than for other satellites of the Saturn system to keep track of their spatial motion.

\section{The Anti-Alignment}
\label{Sec:antialignment}

\cite{Goldreich2003_origin} showed that Prometheus and Pandora have a close approach every 6.2 years when an anti-alignment of their orbits apses happens. Furthermore, they showed that these events, which occurred approximately 0.61, 6.9, 13.1, and 19.3 years after Voyager II closest approach to Saturn in August 1981, are one of the sources of their chaotic motion.

In \autoref{fig:foi_lag_all}, we notice that the difference between the models remained smaller than $0.2\degr$ and started to grow after six years from the initial epoch, in contrast with the Lyapunov time of about 3.3 years reported by \cite{Cooper2004}. In other words, we found regions where the trajectory is well-behaved for twice the Lyapunov time. Probably, the reason is that we started the integration close to the 1982 anti-alignment while the model diverged near the 1988 anti-alignment. In comparison, the Lyapunov exponent is calculated for a period long enough to cross many anti-alignments.

We performed simulations using initial conditions on the epochs of anti-alignment. The intention is to minimise the difference to SAT393 of this specific event and maximise the interval without another close encounter. We also included the anti-alignments at 25.5, 31.7, and 37.9 years after the referred epoch and an integration starting at the symbolic epoch J2000.

In each of these simulations, the integration was done backward up to 1980 and forwarded up to 2020 with an output every day. All the complete set of constants shown in \autoref{tab:constants-sat393} was used.

\begin{figure}
    \centering
    \includegraphics[width=\linewidth]{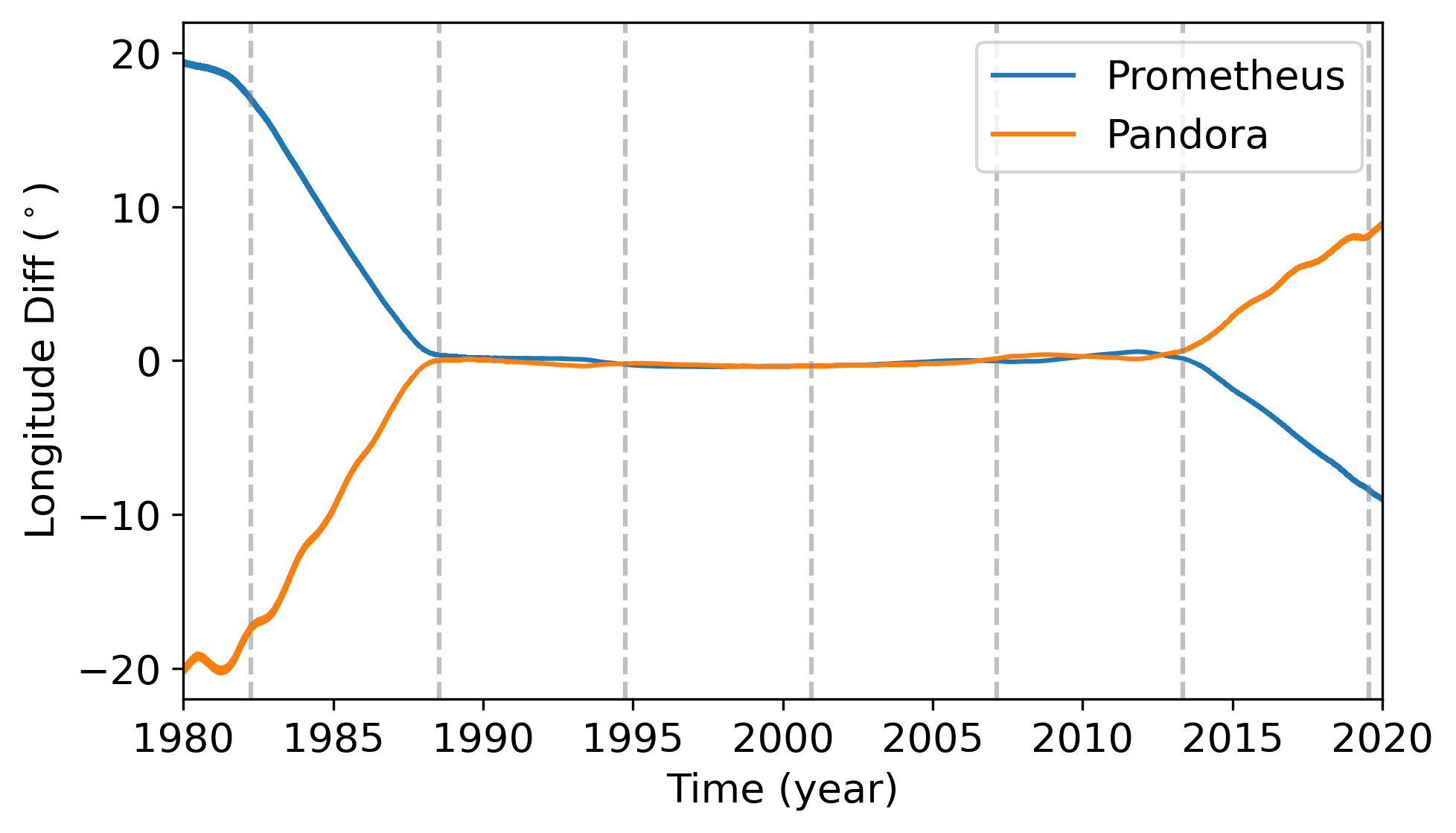}
    \caption{Difference in true longitude between BOSS and SAT393 starting an integration forward and backward in time during the anti-alignment of December 2000. Each vertical dashed line marks an anti-alignment between the orbits of Prometheus and Pandora.}
    \label{fig:anti-alignment}
\end{figure}

\autoref{fig:anti-alignment} shows an integration started in the anti-alignment of December 2000. We note that the first anti-alignment before and after the initial epoch, respectively in 1994 and 2007, does not produce any significant divergence between our model and SAT393. Only when close to the second closest approach, in 1988 and 2013, the model started to diverge. Between these epochs, the difference grows linearly, showing only a small difference in the semi-major axis.

In appendix \ref{App:anti-alignment}, we present the difference between BOSS and SAT393 from integrations starting at different anti-alignments. We note there is no visual pattern between these tests. However, in most cases, we can still recover the orbits of Prometheus and Pandora between the starting epoch and the subsequent anti-alignment. For the case shown in \autoref{fig:anti-alignment}, the trajectories are recovered for almost two close encounters.

As stated in \autoref{Sec:introduction}, we are most interested in identifying the main contributors to the chaotic dynamical evolution of Prometheus and Pandora beyond themselves. It is clear from \autoref{fig:anti-alignment}, and also from \autoref{fig:fitted_to_jpl}, that, to achieve our goals, we should analyse the orbits between two consecutive anti-alignments, where the chaotic motion caused by a close encounter is reduced.

\section{The Relevant Satellites}
\label{Sec:relevant-satellites}

\cite{Cooper2004} already have studied the dynamical influence of the Saturnian satellites. However, although they notice the importance of Titan and Janus-Epimetheus, they do not discuss the significance of the other satellites on the orbits of Prometheus and Pandora.

Our final analysis consisted on verifying the objects that most affect Prometheus and Pandora's orbits in an interval between two consecutive closest approaches. For this, we built an experiment to emphasise the direct perturbation of each test object: first, we integrate the orbits of Prometheus and Pandora using the complete set of objects present in \autoref{tab:constants-sat393} as a base model; then an integration is made removing the test object from the list; finally, the difference in true longitude to the base model is computed.

In this experiment, the direct influence of the tested perturbing object does not affect Prometheus and Pandora, as the object is removed from the system. However, as explained in \autoref{Sec:methodology}, the orbits of the remaining perturbing objects come directly from the SAT393 model, which in turn were determined considering the object we are removing. Thus, the indirect perturbation of the removed object is still preserved in our simulations.

We verified the dynamical perturbations on Prometheus and Pandora caused by all the bodies present in SAT393 whose masses are known: the Sun, the Giant Planets, Enceladus, Tethys, Dione, Rhea, Titan, Mimas, the pair Janus-Epimetheus, Helene, Atlas, Hyperion, Iapetus, and Phoebe. These integrations were done in an interval of 5 years between two consecutive anti-alignment in four distinct periods covering the epochs of the ring plane crossing and the Cassini observations: 1995-2000, 2001-2006, 2007-2012, and 2013-2018. As a basis for comparison, we also integrated the orbit of Prometheus without the direct perturbation of Pandora and vice-versa

\begin{figure*}
    \centering
    \includegraphics[width=\linewidth]{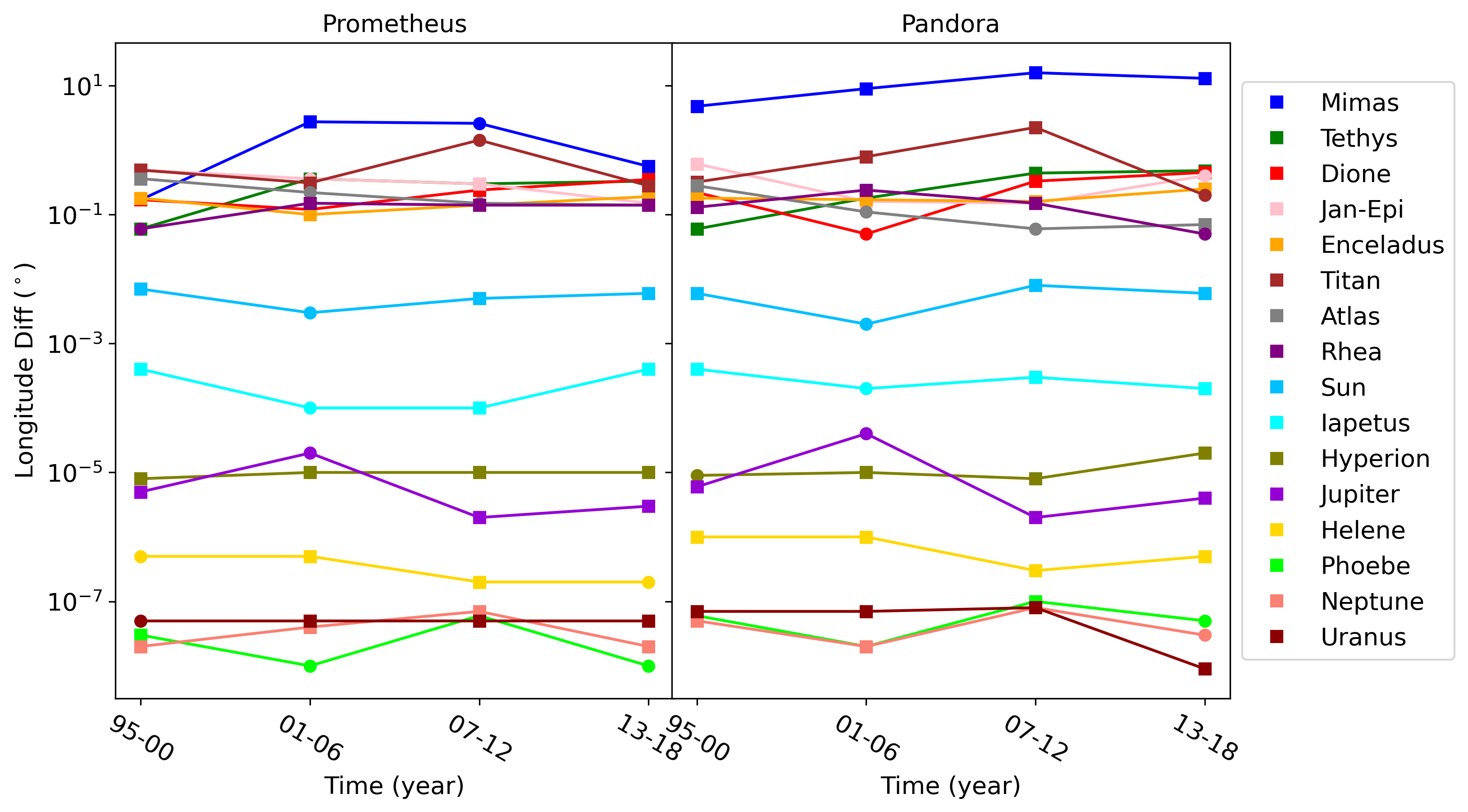}
    \caption{Visual representation of \autoref{tab:maximum-difference}. Each point is the maximum absolute difference in the true longitude of the orbits of Prometheus and Pandora caused by the presence of a specific perturbing object. Square points represent the negative values (lag), while the dots represent the positive ones (ahead).}
    \label{fig:maximum-difference}
\end{figure*}

\autoref{tab:maximum-difference} and \autoref{fig:maximum-difference} show the maximum difference in true longitude for Prometheus and Pandora caused by the exclusion of each test object in each period considered. The difference is in the sense model with minus model without the object, i.e., base model minus model without the object.

\begin{table}
    \centering
    \caption{Maximum difference in true longitude in the orbits of the Saturnian satellites Prometheus and Pandora when the direct perturbation of a body is ignored. The variations were computed in intervals between two consecutive anti-alignments.}
    \begin{tabular}{l|c|c|c|c|c}
        Perturber & Sat.& 1995-2000 & 2001-2006 & 2007-2012 & 2013-2018 \\ \hline \hline
        \multirow{2}{*}{Mimas} & Pro & +0.17 & -2.75 & +2.60 & -0.56 \\
            & Pan & -4.79 & -8.98 & -15.85 & -12.95 \\ \hline
        Prometheus & Pan & 2.69 & 0.28 & -4.47 & 2.48 \\ \hline
        Pandora & Pro & -1.98 & -0.33 & 3.62 & -1.70 \\ \hline
        \multirow{2}{*}{Enceladus} & Pro & -0.18 & -0.10 & -0.14 & -0.19 \\
            & Pan & -0.18 & -0.17 & -0.16 & -0.25 \\ \hline
        \multirow{2}{*}{Tethys} & Pro & -0.06 & -0.36 & -0.30 & -0.33 \\
            & Pan & -0.06 & -0.18 & -0.44 & -0.48 \\ \hline
        \multirow{2}{*}{Dione} & Pro & -0.17 & -0.12 & -0.24 & -0.35 \\
            & Pan & +0.22 & +0.05 & -0.33 & -0.45 \\ \hline
        \multirow{2}{*}{Rhea} & Pro & +0.06 & -0.15 & -0.14 & -0.14 \\
            & Pan & -0.13 & -0.24 & -0.15 & +0.05 \\ \hline
        \multirow{2}{*}{Titan} & Pro & -0.49 & -0.31 & +1.43 & -0.28 \\
            & Pan & -0.32 & -0.79 & -2.25 & +0.20 \\ \hline
        Janus- & Pro & -0.49 & +0.36 & -0.30 & +0.15 \\
        Epimetheus  & Pan & -0.61 & -0.16 & -0.15 & +0.40 \\ \hline
        \multirow{2}{*}{Atlas} & Pro & -0.36 & +0.22 & +0.15 & -0.14 \\
            & Pan & -0.28 & +0.11 & +0.06 & -0.07 \\ \hline
        \multirow{2}{*}{Hyperion} & Pro & $-8\times 10^{-6}$ & $-1\times 10^{-5}$ & $-1\times 10^{-5}$ & $-1\times 10^{-5}$ \\
            & Pan & $+9\times 10^{-6}$ & $-1\times 10^{-5}$ & $-8\times 10^{-6}$ & $-2\times 10^{-5}$ \\ \hline
        \multirow{2}{*}{Iapetus} & Pro & $-4\times 10^{-4}$ & $+1\times 10^{-4}$ & $-1\times 10^{-4}$ & $-4\times 10^{-4}$ \\
            & Pan & $-4\times 10^{-4}$ & $+2\times 10^{-4}$ & $-3\times 10^{-4}$ & $-2\times 10^{-4}$ \\ \hline
        \multirow{2}{*}{Phoebe} & Pro & $-3\times 10^{-8}$ & $+1\times 10^{-8}$ & $-6\times 10^{-8}$ & $+1\times 10^{-8}$ \\
            & Pan & $-6\times 10^{-8}$ & $+2\times 10^{-8}$ & $-1\times 10^{-7}$ & $+5\times 10^{-8}$ \\ \hline
        \multirow{2}{*}{Helene} & Pro & $+5\times 10^{-7}$ & $-5\times 10^{-7}$ & $-2\times 10^{-7}$ & $+2\times 10^{-7}$ \\
            & Pan & $-1\times 10^{-6}$ & $-1\times 10^{-6}$ & $-3\times 10^{-7}$ & $-5\times 10^{-7}$ \\ \hline
        \multirow{2}{*}{Sun} & Pro & $-7\times 10^{-3}$ & $+3\times 10^{-3}$ & $-5\times 10^{-3}$ & $-6\times 10^{-3}$ \\
           & Pan & $-6\times 10^{-3}$ & $+2\times 10^{-3}$ & $-8\times 10^{-3}$ & $-6\times 10^{-3}$ \\ \hline
        \multirow{2}{*}{Jupiter} & Pro & $-5\times 10^{-6}$ & $+2\times 10^{-5}$ & $-2\times 10^{-6}$ & $-3\times 10^{-6}$ \\
          & Pan & $-6\times 10^{-6}$ & $+4\times 10^{-5}$ & $-2\times 10^{-6}$ & $-4\times 10^{-6}$ \\ \hline
        \multirow{2}{*}{Uranus} & Pro & $+5\times 10^{-8}$ & $-5\times 10^{-8}$ & $-5\times 10^{-8}$ & $-5\times 10^{-8}$ \\
          & Pan & $-7\times 10^{-8}$ & $-7\times 10^{-8}$ & $-8\times 10^{-8}$ & $-9\times 10^{-9}$ \\ \hline
        \multirow{2}{*}{Neptune} & Pro & $-2\times 10^{-8}$ & $-4\times 10^{-8}$ & $-7\times 10^{-8}$ & $-2\times 10^{-8}$ \\
          & Pan & $-5\times 10^{-8}$ & $-2\times 10^{-8}$ & $-8\times 10^{-8}$ & $+3\times 10^{-8}$ \\ \hline
    \end{tabular}
    \label{tab:maximum-difference}
\end{table}

First of all, it is clear that the planets, the Sun, and the satellites Hyperion, Iapetus, Phoebe, and Helene are not significant on the orbits of Prometheus and Pandora. This is expected since they are the smallest or the most distant objects from these satellites. 

For the case of Mimas, it is known that the near 3:2 resonance with Pandora is of major importance \citep{Evans2001}. In fact, in our tests, the difference in longitude of Pandora caused by Mimas was always more prominent than that caused by Prometheus reaching $\Delta\lambda \sim 15\degr$ in the interval 2007-2012. The difference in the longitude of Prometheus' orbit caused by Mimas was smaller than $3\degr$ ($\Delta\lambda \sim 5\degr$) and was more significant than the perturbation caused by Pandora only in the interval 2001-2006.

\begin{figure*}
    \centering
    \includegraphics[width=\linewidth]{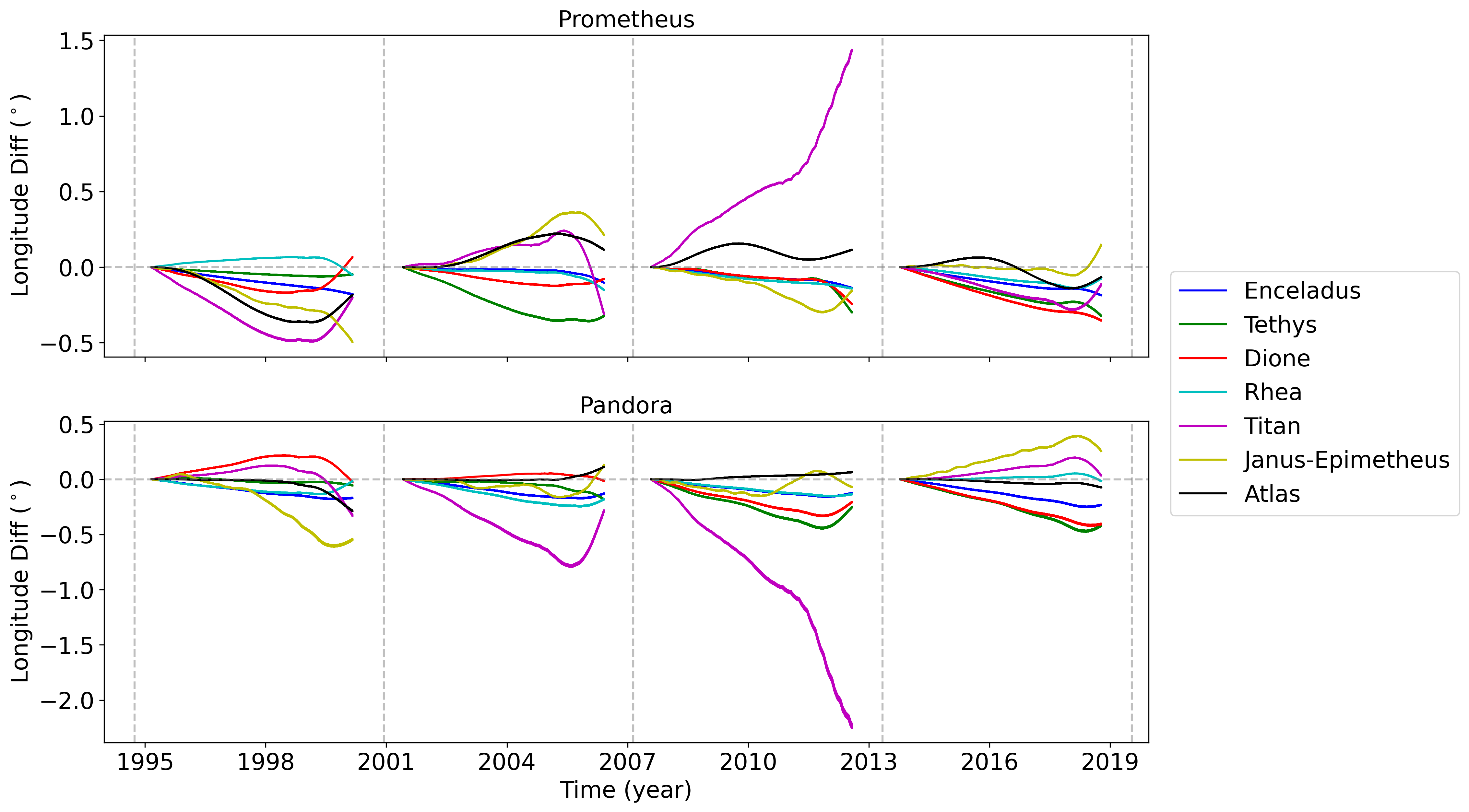}
    \caption{Difference in true longitude between BOSS and SAT393 for integrations where a specific perturber was removed compared to the one where all perturbers were considered. The variation caused by the absence of Mimas is not shown as it is much larger than the others.}
    \label{fig:dynamical_influence}
\end{figure*}

For the remaining satellites, their dynamical influences are of the foremost importance but much less significant than Mimas. The variation over time can be seen in \autoref{fig:dynamical_influence} and the maximum difference reached in each interval is summarised in \autoref{tab:maximum-difference}.

First, we notice that, in the majority of the cases, the variation in true longitude in Prometheus and Pandora caused by the missing of a specific perturber is in the opposite direction for each other. It is interesting to note that, while the missing of Mimas caused a lag in the orbit of Prometheus in 2001-2006, Prometheus would be ahead of its expected position in 2007-2012 by almost the same amount.

For Enceladus, the absence of its direct perturbation caused a lag of $\sim 0.2\degr$ in both satellites regardless of the epoch studied. Without Tethys, almost no lag was identified in 1995-2000, but a lag of nearly $0.5\degr$ was identified in 2013-2018. For Dione, the period 2013-2018 is also when it would cause the most considerable lag ($\sim 0.4\degr$) in the orbits of Prometheus and Pandora. For Rhea, the lags are consistent in all the intervals with variations smaller than $0.25\degr$.

The missing of the direct perturbation of Titan caused a lag of $2.25\degr$ in the orbit of Pandora, and a lead of $1.43\degr$ in the orbit of Prometheus, in the interval 2007-2012. During this period, only Mimas was more important than Titan. In the interval 2001-2006, the perturbation of Titan on Pandora was also more significant than that caused by Prometheus. However, in the interval 2013-2018, the direct perturbation of Titan is less important than Mimas, Tethys, and Dione.

The pair Janus-Epimetheus also contributes to Prometheus' and Pandora's orbital evolution. \cite{Cooper2004} showed that when Epimetheus switches its orbit with Janus, the 17:15 eccentricity resonance sweeps across Prometheus' orbit. At the same time, the 21:19 eccentricity resonance sweeps across Pandora's orbit. The direct perturbation of Janus-Epimetheus was found to be the second most crucial interaction with Prometheus and Pandora in the period 1995-2000 ($\Delta\lambda\sim 0.6\degr$), but similar to the perturbation by other objects in the remaining periods tested.

Atlas has also been identified as having mean motion resonances 54:53 with Prometheus and 70:67 with Pandora by \cite{Spitale2006}. Thus, in our simulations, the direct perturbation of Atlas affects Prometheus more than Pandora. However, in the periods tested, the perturbation of Atlas is not higher than that caused by other satellites.

\section{Final Comments}
\label{Sec:final-comments}

The dynamics of the orbits of Prometheus and Pandora is very complex. After the discovery, their orbits presented longitudinal lags from the expected positions as observed by the Hubble Space Telescope during the Ring Plane Crossing \citep{Evans2001}. Furthermore, a chaotic behaviour associated with a 121:118 mean motion resonance was found responsible for the lags as described by \cite{Goldreich2003_chaotic}. Because of it, an anti-alignment of the orbits of these satellites happens every 6.2 years, amplifying their chaotic dynamical evolution. However, as pointed out by \cite{deSantana2020}, the mutual interaction between Prometheus and Pandora alone is not enough to explain the observed lags. In this regard, we realised a study of the orbits of Prometheus and Pandora in an attempt to identify the bodies that could potentially contribute to their chaotic motion.

To study the period covered by the observations from Voyager I (1980) until Cassini (2016), we used the JPL SAT393 ephemeris as a comparative. In an attempt to replicate the SAT393 ephemeris, using the same dynamical parameters and the ephemeris itself as initial conditions, our model quickly diverged from the ephemeris. At the same time, our model managed to replicate the orbits of Atlas, Titan, Janus-Epimetheus, and Phoebe. Unfortunately, we could not reliably propagate the orbits of Prometheus and Pandora for more than a few years, even when we forced our model to be the same as the SAT393 ephemeris. This shows how sensitive their orbits are.

Our model only diverged from the SAT393 one once Prometheus' and Pandora's orbits were close to the anti-alignment. To minimise the effects of a specific anti-alignment and maximise the interval without experiencing another one, we made simulations starting at these epochs. We notice that the difference between the models started when crossing to the first anti-alignment. In one particular case, Dec. 2000, shown in \autoref{fig:anti-alignment} only the second anti-alignment was able to cause a significant divergence between our model and the SAT393 while the difference in true longitude between two consecutive anti-alignments grows linearly, meaning a difference in the semi-major axis. As expected, the anti-alignment between Prometheus and Pandora is the leading cause of the instability of their orbits. We were able to replicate the SAT393 model for a period of $\sim 6$ years in almost all our tests, compared to the Lyapunov time of 3.3 years.

Finally, to identify the essential satellites to reproduce the motions of Prometheus and Pandora, we made numerical integrations between two consecutive anti-alignments. In this interval, we minimise the importance of the closest encounters, which better show us other effects. Furthermore, the tests were made comparing a simulation without a specific satellite and an orbit where all satellites were present, thus considering only the direct perturbation of the satellites.

These simulations showed us that Mimas, Enceladus, Tethys, Dione, Rhea, Titan, Janus-Epimetheus, and Atlas seemed to be of utmost importance for the orbits of Prometheus and Pandora. All of them could cause a variation in the longitude of Prometheus' and Pandora's orbits of about $0.2\degr$, which will be amplified when crossing an anti-alignment. Thus, we conclude that any model that intends to study the trajectories of Prometheus and Pandora should consider at least these nine satellites.

\cite{Quillen2014} analysed the three-body resonances in the Uranian satellite system in an attempt to identify theoretically and numerically which types of three-body resonances might be important. Our work was focused on the numerical simulations of the Prometheus' and Pandora's orbits in the context of the observations from Voyager I, in 1980, up to Cassini, in 2016. Thus we do not discuss unknown resonances that could explain the results obtained. However, we encourage further analysis to look for possible three-body resonances in the system, in the view of a large number of two-body resonances associated with both satellites: Prometheus-Pandora 121:118 mean motion resonance; Mimas-Pandora 3:2 corotation resonance, which in turn, it is known that Mimas is in a 2:1 mean motion resonance with Tethys; Epimetheus-Prometheus and Epimetheus-Pandora respectively 17:15 and 21:19 eccentricity resonances, Epimetheus is in corotation with Janus; Atlas-Prometheus and Atlas-Pandora respectively 54:53 and 70:67 mean motion resonances.

\section*{Acknowledgements}\label{acknowledgements}
The authors acknowledge the respective grants:
FAPESP (proc. 2016/24561-0 and 2018/11239-8), CNPq (proc. 305210/2018-1), CAPES and DFG German Research Foundation project 446102036;

\section*{Data Availability}

No new data were generated or analysed in support of this research.



\bibliographystyle{mnras}
\bibliography{main} 




\appendix

\section{Graphics of Anti-Alignment}
\label{App:anti-alignment}

In \autoref{Sec:antialignment} we showed how the anti-alignment between Prometheus' and Pandora's orbits, which happens every 6.2 years, is of the foremost importance. Thus, we computed their orbits starting at different anti-alignments to minimise the difference between our model and the JPL SAT393 at this specific configuration and maximise the time before the subsequent close encounter.

Here we present the graphics showing the difference between BOSS and JPL SAT393 starting at the anti-alignments that occurred after 0.61, 6.9, 13.1, 25.5, 31.7, and 37.9 years after Voyager II closest approach to Saturn. These epochs are respectively: 1982-Apr, 1988-Jul, 1994-Set, 2007-Feb, 2013-May, 2019-Jul. The anti-alignment that happened 19.3 years after the referred epoch, in 2000-Dec, is presented in \autoref{fig:anti-alignment}.

\begin{figure*}
\center
\subfigure[ref1][1982-Apr]{\includegraphics[width=0.49\textwidth]{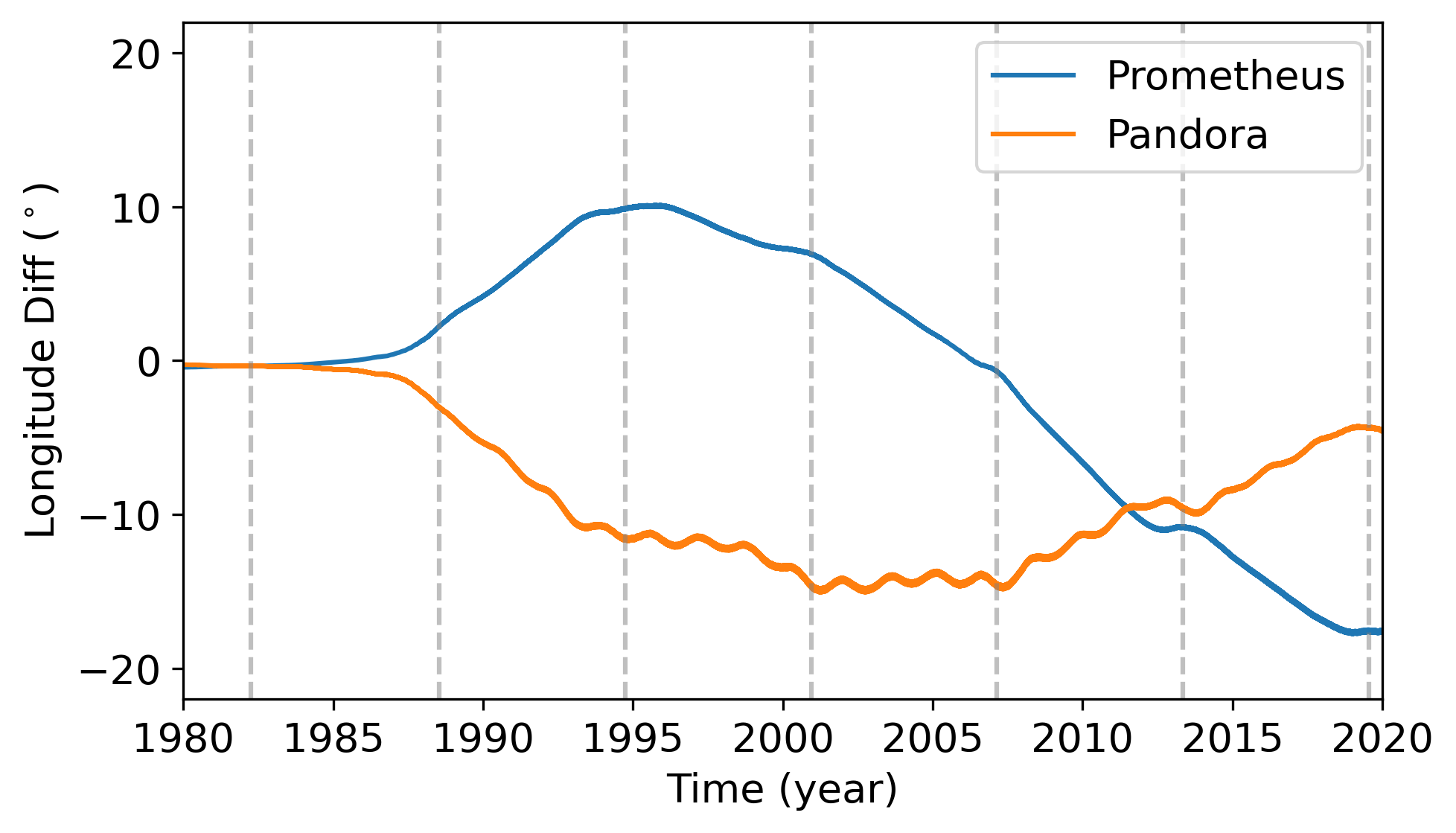}}
\subfigure[ref2][1988-Jul]{\includegraphics[width=0.49\textwidth]{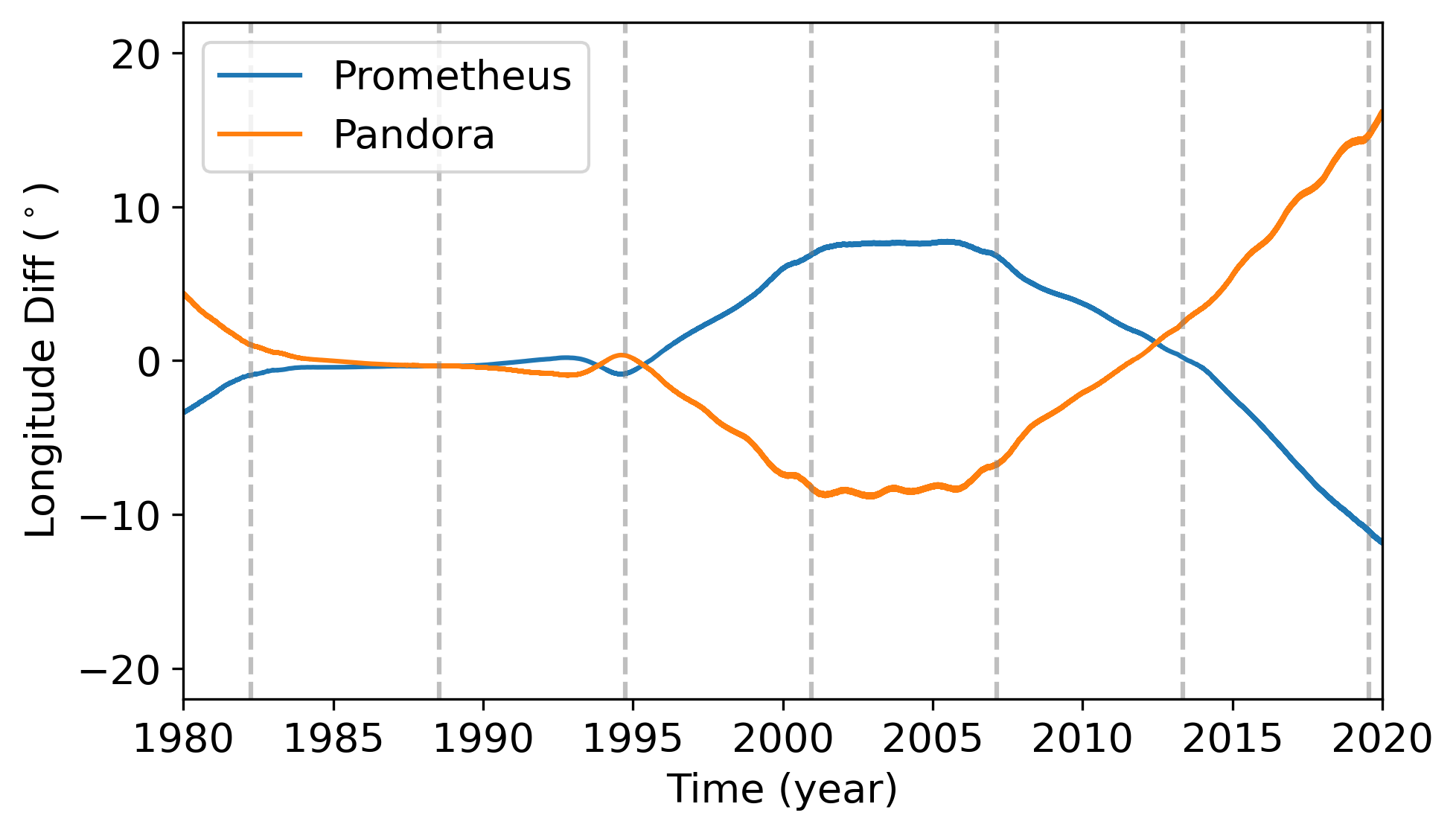}}
\subfigure[ref3][1994-Set]{\includegraphics[width=0.49\textwidth]{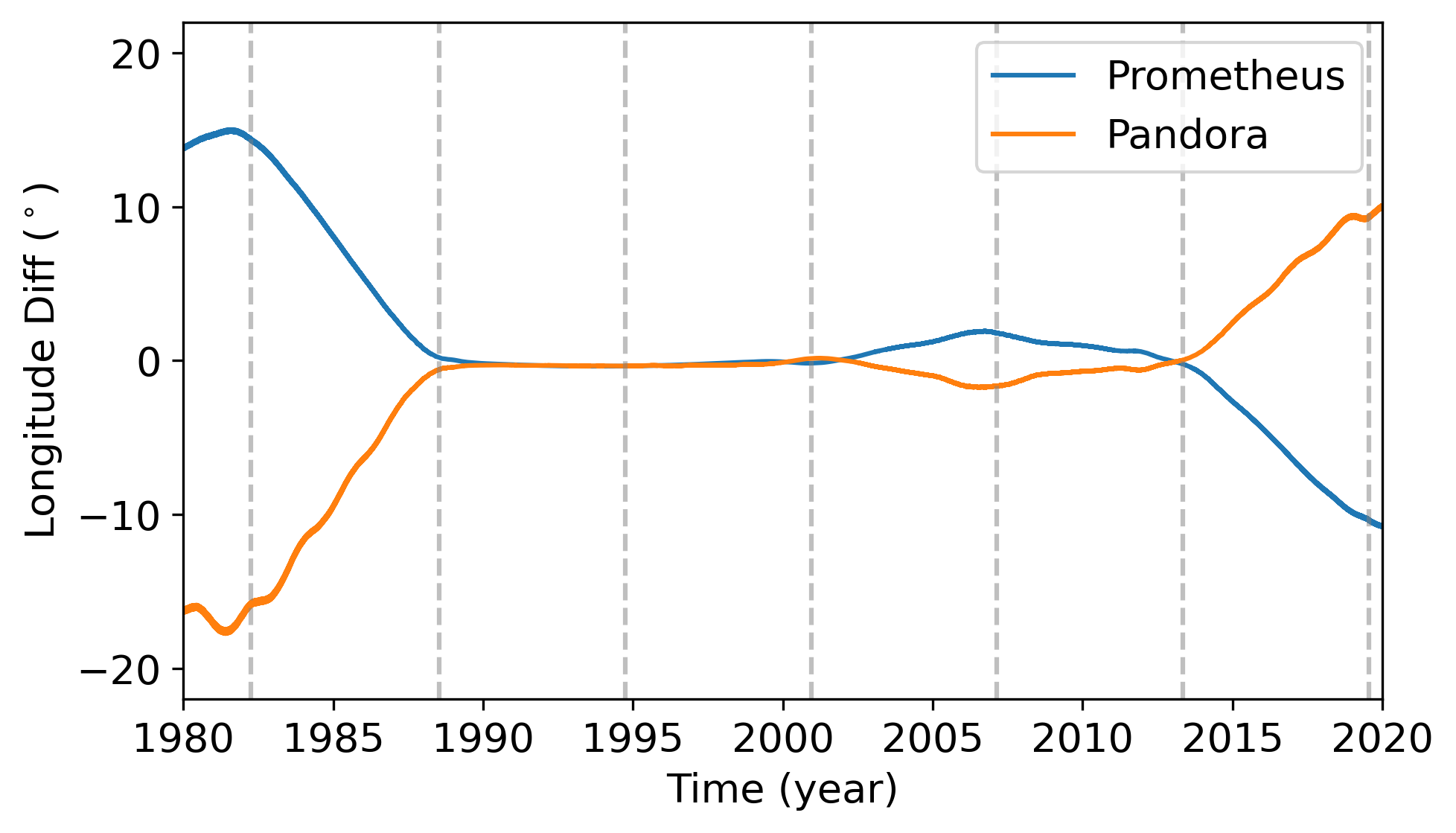}}
\subfigure[ref4][2007-Feb]{\includegraphics[width=0.49\textwidth]{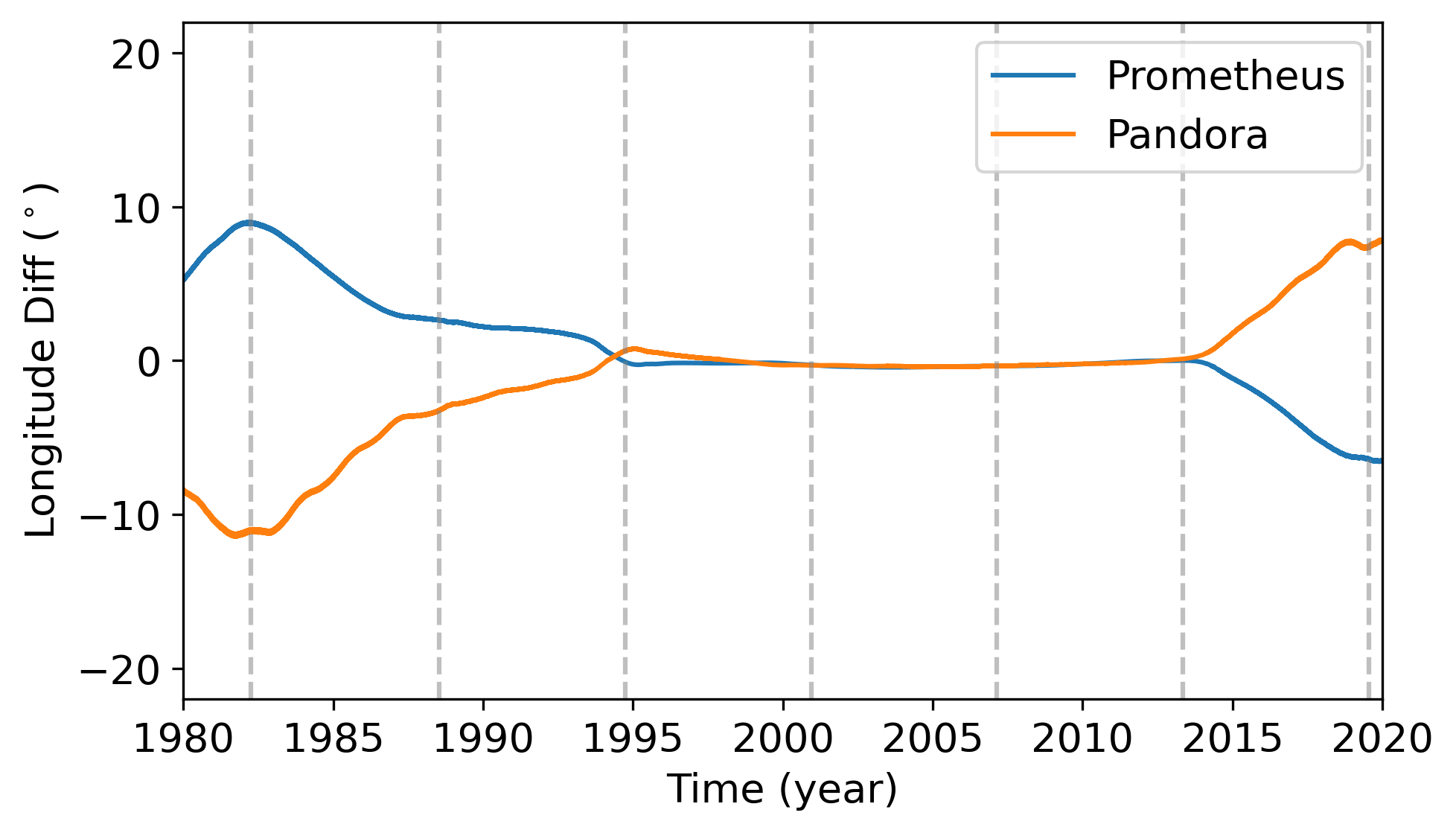}}
\subfigure[ref5][2013-May]{\includegraphics[width=0.49\textwidth]{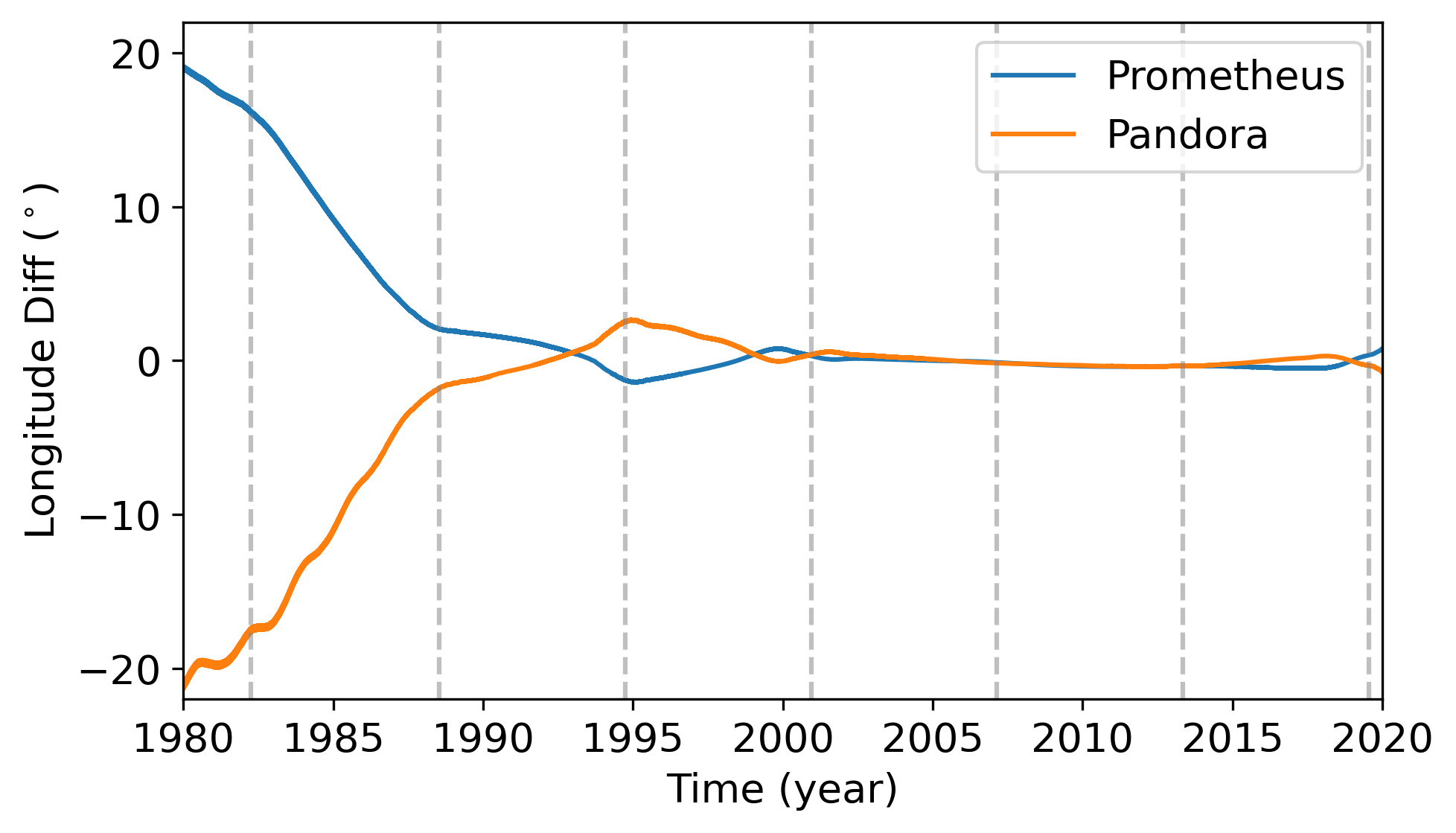}}
\subfigure[ref6][2019-Jul]{\includegraphics[width=0.49\textwidth]{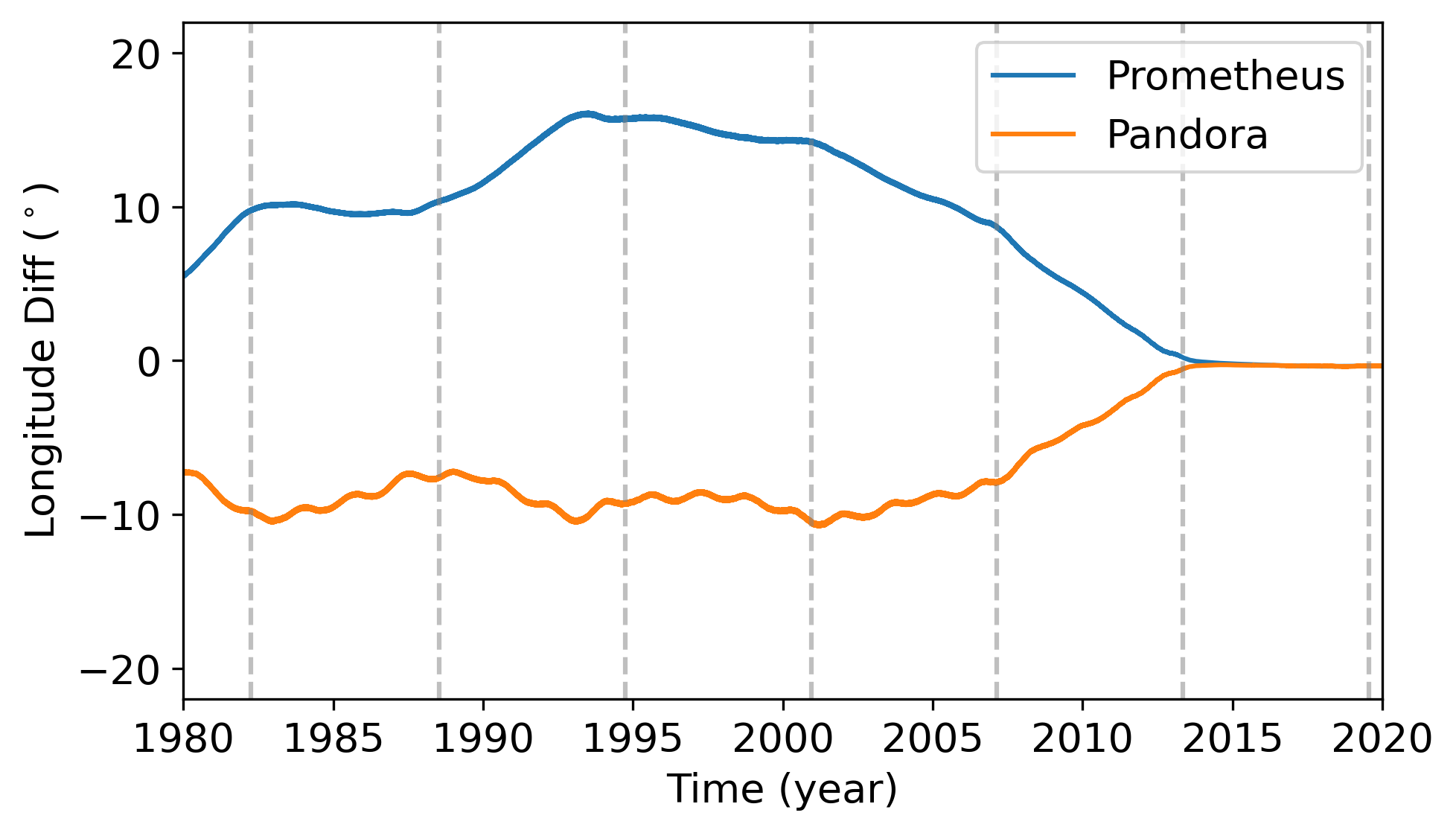}}
\caption{Difference in true longitude between BOSS and SAT393 for integrations starting at different anti-alignment epochs. \label{fig:annex_dynamical_influence}}
\end{figure*}



\bsp	
\label{lastpage}
\end{document}